# Probing the Electrical Properties of Overlapped Graphene Grain Boundaries by Raman spectroscopy

Rahul Rao,* Neal Pierce, Avetik R. Harutyunyan

Honda Research Institute USA, Columbus, OH, USA 43212

Correspondence – rrao@honda-ri.com

**Abtract**

The effect of grain boundaries and wrinkles on the electrical properties of polycrystalline graphene is pronounced.  Here we investigate the stitching between grains of polycrystalline graphene, specifically, overlapping of layers at the boundaries, grown by chemical vapor deposition (CVD) and subsequently doped by the oxidized Cu substrate. We analyze overlapped regions between 60 – 220 nm wide via Raman spectroscopy, and find that some of these overlapped boundaries contain AB–stacked bilayers. The Raman spectra from the overlapped grain boundaries are distinctly different from bilayer graphene and exhibit splitting of the *G* band peak. The degree of splitting, peak widths, as well as peak intensities depend on the width of the overlap. We attribute these features to inhomogeneous doping by charge carriers (holes) across the overlapped regions via the oxidized Cu substrate. As a result, the Fermi level at the overlapped grain boundaries lies between 0.3 and 0.4 eV below the charge neutrality point. Our results suggest an enhancement of electrical conductivity across overlapped grain boundaries, similar to previously observed measurements[1].  The dependence of charge distribution on the width of overlapping of grain boundaries may have strong implications for the growth of large-area graphene with enhanced conductivity.

1. Introduction

The continuity of polycrystalline graphene and the minimization of its grain boundaries (GBs) remains a big challenge towards its integration into large-scale applications [2-5]. Recent reports indicate that the stitching, or connection between grains in graphene grown by CVD has a significant impact on the electrical transport across the GBs [1,6-9] as well as the mechanical integrity of graphene films [10]. Moreover, GBs in graphene are known to be high-reactivity sites[11], and have been observed to induce a higher degree of oxidation of the underlying copper substrates [12-14]. As a result, much effort has been directed towards the growth of large area graphene films by the suppression of nucleation sites [4,15-18].

At the atomic level the interconnection between graphene grains in CVD-grown graphene has been shown to comprise of alternating pentagons and heptagons along the seams [19,20]. However, in addition to atomic bonding between graphene grains, graphene grains have also been observed to overlap each other when the grain growth rate is sufficiently low. Overlapped grains up to 1 μm in width have been observed [1,21]. Surprisingly, the electrical transport across the overlapped GB can be better than the atomically interconnected (and disordered) grain boundary, with the electrical conductance an order of magnitude higher [1]. This result from Ref. 1 suggests that the creation of GBs with engineered widths would potentially be an exciting advance towards achieving continuous electrical conductivity across large-area polycrystalline graphene films. Although the atomic structure of these overlapped GBs is not fully known, evidence for misoriented stacking between the overlapped layers in few layer graphene samples has been found by high-resolution transmission electron microscopy (HRTEM) [22]. In addition to graphene, overlapped grains have also recently been observed in CVD-grown $MoS_2$ [23,24], although their impact on properties is not known. Overlaps between 2D layers during growth could be universal, and the possibility of electrically conductive overlaps across narrow atomic layers hints at the prospect of utilizing these nanoribbons in a variety of exciting applications.

Here we investigate the electronic properties of a particular set of overlapped GBs between graphene grains on oxidized Cu substrates by Raman spectroscopy. GBs with overlap widths varying between 60 and 220 nm are first mapped by scanning electron microscopy (SEM) and Raman spectra collected from the GBs as well as adjacent graphene grains for comparison. In our doped samples a small percentage of overlapped GBs (<10%) exhibit Raman spectra similar to that of bilayer graphene (BLG) with one difference from pristine BLG - the Raman *G* band from the overlapped regions splits into two components, and the degree of splitting, peak widths, as well as peak intensities depend on the width of the overlapped layer. These changes in the Raman spectra from the GBs are attributed to inhomogeneous doping by the oxidized Cu substrate, with doping levels increasing with the GB overlap width, consequently tuning the Fermi levels across the GBs between -0.3 and -0.42 eV. Our results suggest that engineered overlapping of GBs offer the potential to exploit them in a variety of devices involving BLG nanoribbons [25], as well as unique device geometries such as nanoribbon diodes with negative differential resistance [26] and electromechanical switches based on overlapping nanoribbons [27].

## 2. Experimental Methods

Graphene growth on Cu foil is performed by atmospheric pressure CVD. The Cu foil substrates (15 µm thick) are loaded into the center of a tubular (2" diameter) quartz furnace and heated from room temperature to a growth temperature of 1020 °C with a ramp rate of 50 °C/min. Once the growth temperature is reached, it is held for 60 minutes to anneal the Cu foils. Both the ramp and the annealing are performed under a constant flow of argon (550 sccm) and hydrogen (30 sccm). Graphene growth starts upon the introduction of $CH_4$ (20 sccm of 1000 ppm $CH_4$ in Ar) along with the $Ar/H_2$ gases. The growth time ranges between 30 to 60 minutes. Following growth the samples are allowed to cool down to room temperature naturally under $Ar/H_2$.

Sample characterization is performed by scanning electron microscopy (SEM, Zeiss Ultra), conductive atomic force microscopy (cAFM), and Raman spectroscopy (Renishaw inVia Raman microscope). Raman spectra are collected from the GBs with two laser excitations ($E_{laser}$ = 1.96, 2.33, and 2.41 eV) at low laser powers (~1 mW) in order to avoid heating.

## 3. Results and Discussion

The graphene samples are prepared by atmospheric pressure CVD, which typically produces hexagonal graphene grains on Cu substrates at short growth times. We keep growth times between 30 and 60 min., which are long enough to allow the nucleation and growth of multiple grains such that they merge with each other, but short enough to prevent full coverage of graphene on the substrate. In order to easily locate the graphene grains under an optical microscope for Raman spectroscopy studies, after growth the samples are subjected to mild oxidation for 2 min. at ~180 °C [12,28]. The oxidation process serves a dual purpose – to make the graphene grains visible, and to induce doping [29] of the graphene grains as a means to probe the charge environment across the GBs. Our Raman spectroscopy studies are performed on graphene grains directly on Cu substrates rather than after transferring to $SiO_2$ in order to eliminate stresses/tears in the GBs due to the transfer process, as well as disorder and unintentional doping due to polymer residues. Fig. 1a shows a scanning electron microscope (SEM) image of an overlapped GB with a width of ~100 nm between two graphene grains (indicated by the arrow). The faceted structure of one of the grains can be seen clearly in the inset in Fig. 1a. The discoloration and other aberrations observed in the SEM images are due to the oxidized Cu surface.

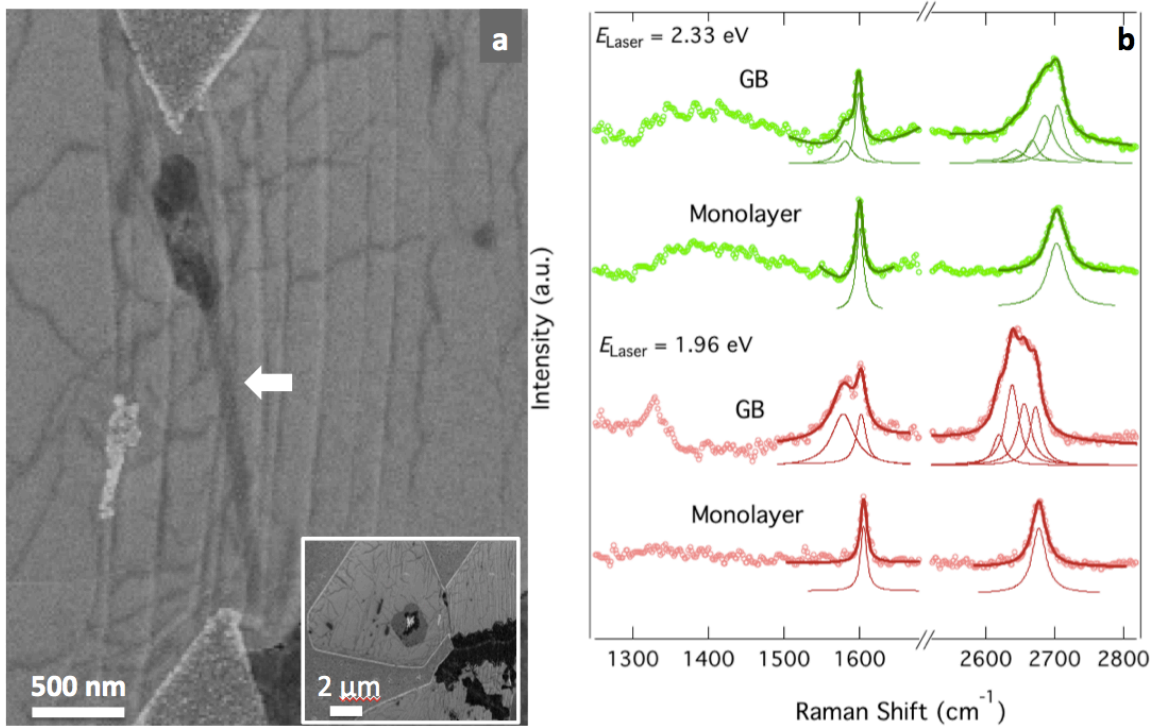

**Fig. 1** – (a) SEM image showing an overlapped grain boundary (GB) between two graphene grains. The overlapped region is indicated by the white arrow. The inset in (a) shows a low magnification view of the two polygonal graphene grains. (b) micro-Raman spectra collected from the GB and graphene region next to the GB using $E_{laser}$ = 1.96 eV (633 nm) and 2.33 eV (532 nm). The $G$ and $G'$ bands in the spectra have been fitted with Lorentzian peaks. All spectra have been offset for clarity.

Micro-Raman spectra (laser spot ~ 1 μm diameter) collected with two excitations ($E_{laser}$ = 1.96 eV and 2.33 eV) show the presence of doping (Fig. 1b). The Raman spectra from graphene next to the GB (labeled "monolayer" in Fig. 1b) exhibit the typical peaks, namely the in-plane vibrational $E_{2g}$ vibrational mode ($G$ band) around 1600 cm$^{-1}$ and the second order TO phonon mode ($G'$ band) around 2700 cm$^{-1}$ [30-32]. The $G'$ peaks from the graphene fit well with a single Lorentzian peak having a linewidth (full width half maximum or FWHM) of ~ 25 cm$^{-1}$ [31,32]. The corresponding $G$ bands are also narrow with linewidths of ~ 10 cm$^{-1}$. However, the frequencies of the $G$ bands are blueshifted from 1585 cm$^{-1}$, which is the frequency corresponding to pristine freestanding graphene [33], to ~1605 cm$^{-1}$. This blueshift is likely due to doping from the oxidized Cu substrate [34]. The *G'* band from the graphene is also blueshifted from its undoped frequency (~2650 cm$^{-1}$) to ~2675 cm$^{-1}$. The blueshift of both the *G* and *G'* bands indicates p-type doping in the graphene on the oxidized Cu substrate [35,36]. Note that the spectra shown in Fig. 1b have been baseline corrected to account for the broad luminescence background from the Cu substrate. Another indicator of doping in the graphene is the low intensity ratio between the *G'* and *G* bands ($I_{G'}/I_G$) [34], which is closer to 1, in contrast with the expected value of >5 as observed in pristine graphene on Cu [32]. A low $I_{G'}/I_G$ in graphene is also expected due to doping. A third dispersive peak which is typically observed due to disorder at ~1330 cm$^{-1}$ (~1350 cm$^{-1}$) with $E_{laser}$ = 1.96 eV (with $E_{laser}$ = 2.33 eV) [31] cannot be observed in the graphene Raman spectra, indicating a high degree of crystallinity in the graphene grains.

The Raman spectrum from the GB in Fig. 1a is also blueshifted. However, the *G'* peak from the GB is quite different from that of the nearby graphene region - the *G'* peak exhibits a complex lineshape and can be deconvoluted into four Lorentzian peaks instead of a single peak. The four-Lorentzian peak *G'* band lineshape is unique to BLG [30,31], and strongly suggests that the overlapped GB consists of two graphene layers with Bernal (or AB) stacking between the layers. In other words, the two graphene grains are bounded by a narrow BLG nanoribbon. This result is in contrast to the previous HRTEM study that reported stacking faults at the boundaries between graphene layers in a few layer graphene sample [22]. The difference in stacking order across the overlapped region in our sample compared to Ref. 22 could be due to the different growth rates or CVD method employed (the CVD process employed in Ref. 22 produced multi-layer graphene). It is known that atmospheric pressure CVD produces nearly hexagonal graphene grains with straight edges. In the event that the edges of two growing hexagonal grains line up parallel to each other during growth, one grain could overlap another and retain commensurate stacking between the two grains. In our samples the overlapped GBs that exhibit Raman spectra corresponding to AB-stacked BLG are formed such that the edges of the two graphene grains are more or less parallel to each other (see an example in Fig. S1 with related discussion).

We note that while our laser spot is ~ 1 μm in diameter, we are able to collect spectra from GBs whose widths are much lesser than the spot size. Linescans collected across several GBs (see for example Figs. S2 and S3) show the *G'* band evolving from a single peak lineshape to a broadened, redshifted, and multi-peaked lineshape across the GB. In order to gain further insight into the nature of the peaks comprising the *G'* band in the overlapped GBs, we measure the dispersion of their peak frequencies as a function of excitation laser energy. The four peaks in the *G'* band from BLG typically have a dispersion of ~90 cm$^{-1}$/eV for the laser energy range used in this study (1.96 – 2.41 eV) [31,37]. The dispersion of the four deconvoluted peaks within the *G'* band from the BGs ranges from 73 – 80 cm$^{-1}$/eV (Fig. S4) and is consistent with previously reported values for BLG [31,37], though lower than the values for pristine BLG. The lower values of the peak dispersions between the overlapped GBs and pristine BLG could be the result of disorder or strain (discussed further below). Such a lowering of peak dispersion has been observed previously in disordered single-walled carbon nanotubes [38]. We note that the lineshapes of the G' bands in the overlapped GBs do not match the typical lineshapes from BLG on Cu reported in the literature. Indeed the *G'* band from BLG on Cu foil does not typically exhibit shoulders indicating multiple deconvoluted peaks; its lineshape appears more like a single peak, although broader than the *G'* band from monolayer graphene [18].

Our observation of a complex multi-peak G' band from the overlapped GBs, along with the dispersion of peak frequencies that agrees very well with the expected Raman response from BLG, therefore strongly suggests that these overlapped GBs are composed of AB-stacked BLG. We stress that the occurrence of AB-stacked overlapped GBs is not universal. They comprise a small percentage (<10%) of all the GBs (overlapped and continuous) observed in our samples. As expected, the Raman spectra from overlapped grain boundaries with incommensurate stacking exhibit *G* and *G'* bands with single-Lorentzian lineshapes [31]. Overlaps in graphene can also be caused by wrinkles, which are typically formed due to the different thermal contraction rates of the Cu foil and graphene during cool-down after CVD growth. However, the Raman spectra from the overlapped GBs are unlike spectra from folds or wrinkles in the graphene grains that are produced due to the different cooling rate of the underlying Cu substrate (See Fig. S5 in the Supporting Information for an SEM image and Raman spectrum from a wrinkle in the graphene).

In addition to differences in the lineshapes of the *G'* bands between the overlapped GBs and the monolayer regions, another striking difference in the Raman spectra is in the lineshape of the *G* band. The *G* band from the GB is split into two components, a high frequency peak at ~1604 cm$^{-1}$ and a low frequency peak at ~1580 cm$^{-1}$. Henceforth we will refer to the lower and higher frequency peaks as the *G$^-$* and *G$^+$* peaks, respectively. The $G^-$ peak is broader than the $G^+$ peak in Fig. 1b for both laser excitations. Splits in the *G* band in graphene and BLG have been observed previously due to uniaxial strain [39-41] or due to charge transfer (doping) [42-50]. In the case of uniaxial strains (up to 3%) the *G* band splitting is not accompanied by any measurable disorder or defect-induced Raman peak. However, we do observe a *D* band in the Raman spectrum measured with $E_{laser}$ = 1.96 eV from the GB, as can be seen in Fig. 1b. The D band in the spectrum measured with $E_{laser}$ = 2.33 eV cannot be distinguished clearly due to the higher background noise. In addition to the absence of a *D* band, the strain-induced split components of the *G* band should have similar linewidths, albeit broader lineshapes compared to the *G* band from un-strained graphene [40, 41]. The spectra in Fig. 1b clearly show a broadened $G^-$ peak, while the linewidth of the $G^+$ peak is similar to that of the unstrained graphene *G* band. In fact as discussed below, the $G^+$ peak is also observed to sharpen compared to the *G* band from graphene. If we neglect uniaxial strain, the other possibility is biaxial strain, which is more likely to occur during growth when two graphene grains overlap. However, biaxial strain does not cause splitting of the *G* band [51]. Thus the only remaining cause for the splitting of the *G* band from the overlapped GB is due to charge transfer via inhomogeneous doping. Our claim is further supported by the observation of an increase in intensity of the copper oxide Raman peaks at the GB (Fig. S6), suggesting a correlation between the oxide and the unique features in the Raman spectra from the GB.

Previously, evidence for *G* band splitting has been observed for both electrical [45, 50] and chemical doping [46-49, 52, 53]. The splits have been attributed to optical phonon mixing due to breaking of the inversion symmetry caused by inhomogeneous doping [43-45, 50]. The *G* band in BLG arises from doubly degenerate zone-center phonons with $E_{2g}$ (in-phase or symmetric displacement of atoms in the two layers) and $E_u$ (out-of-phase or antisymmetric displacement of the atoms) symmetries. The anti-symmetric $E_u$ mode is not Raman-active. Thus the *G* band in pristine BLG (or homogeneously doped BLG) does not exhibit any splitting and only a single peak is observed in the Raman spectrum. However, the inversion symmetry of BLG is broken due to inhomogeneous doping, leading to a mixing of the symmetric and anti-symmetric modes and splitting of the *G* band [43, 45, 47]. The resulting two peaks are therefore a superposition of the symmetric and anti-symmetric modes. The frequency difference between the two split peaks depends on the charge separation between the top and bottom graphene layer due to the inhomogeneous doping.

As mentioned above, optical phonon mixing in BLG can be observed through both electrical [45, 50] and chemical doping [46-49, 52, 53]. In our samples the oxidized Cu substrate dopes the graphene layers. Since the two layers in the overlapped GB are exposed to different levels of oxidation, one would expect the overlapped GB to be doped inhomogeneously. Furthermore, the edges along the GBs act as charge traps and could also contribute to the different levels of doping in the bottom and top layers across the overlapped GB. We note here that a splitting of the *G* band has also been observed in the case of scrolled graphene [54]. However, the Raman spectrum from the scrolled graphene contains low frequency radial breathing-like modes, which we do not observe in the spectra from the overlapped GBs. We also note that the splitting of the *G* band is not observed from GBs where the graphene grains merged smoothly without any overlaps.

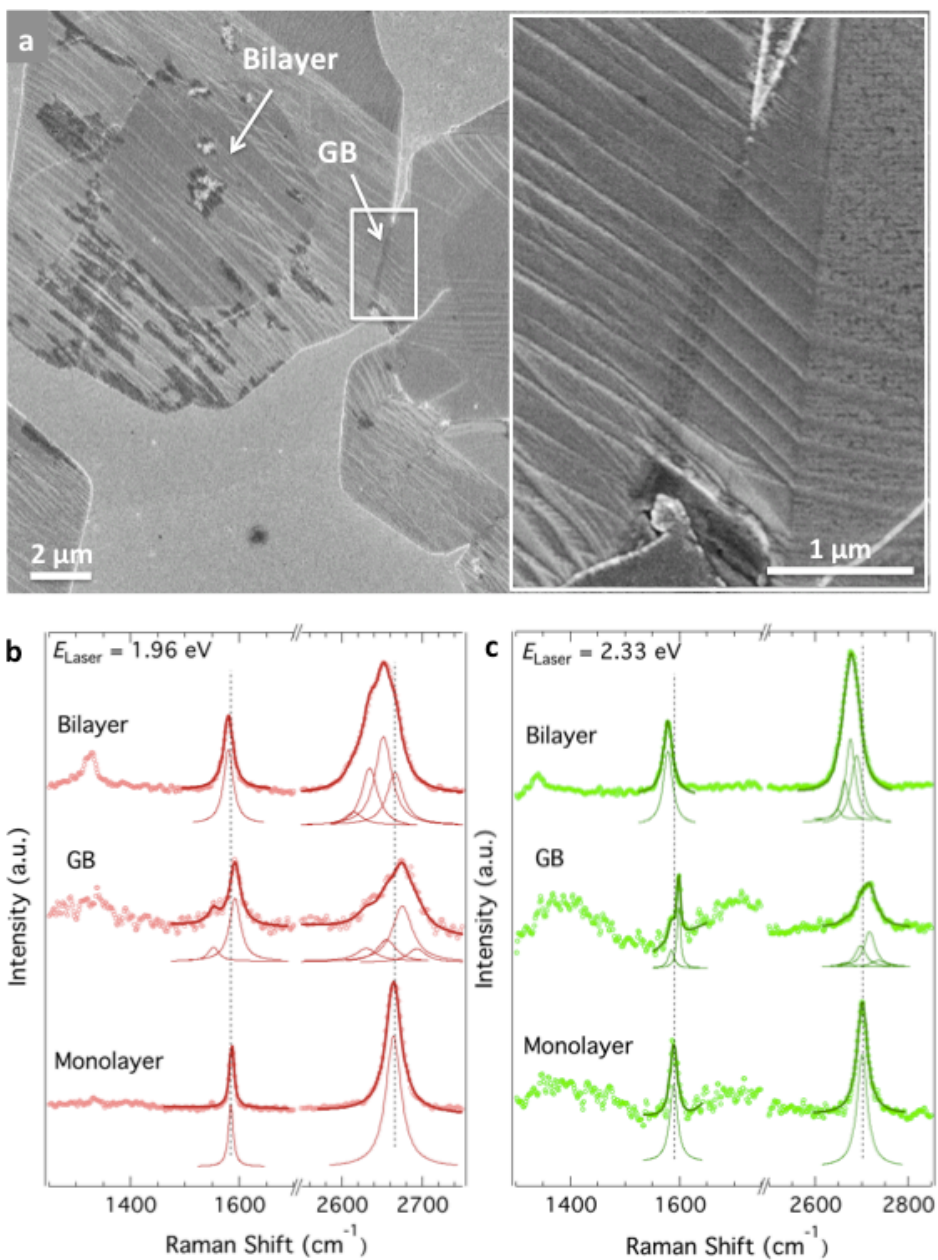

Fig. 2 – (a) SEM image showing two graphene grains with overlapped GB. A BLG grain can be seen in the center of one of the monolayer grains. Inset – high magnification view of the overlapped GB. (b) and (c) Raman spectra collected using, (b) $E_{laser}$ = 1.96 eV and, (c) $E_{laser}$ = 2.33 eV from the monolayer, bilayer and GB regions. The $G$ and $G'$ bands have been fitted with Lorentzian peaks. All spectra have been offset for clarity.

In order to confirm that the *G* band splitting is unique to the overlapped GBs, we measure Raman spectra from BLG regions grown within the graphene grains. During the CVD growth of graphene, a second layer is known to nucleate and grow under the first layer, typically with the same nucleation center [55]. Such an example is shown in Fig. 2a where a hexagonal BLG grain with a darker contrast can be seen in the middle of the larger graphene grain. To the right of the BLG grain lies an overlapped GB, visible in the higher magnification inset in Fig. 2a. The corresponding Raman spectra collected with $E_{laser}$ = 1.96 eV and 2.33 eV from the graphene, GB and bilayer regions are shown in Figs. 2b and 2c. The splitting of the *G* band in the GB is clearly evident and is similar to the spectra shown in Fig. 1b. However, the Raman spectra from the BLG do not exhibit any splitting of the *G* bands. In fact, the *G* band from the BLG appears at a lower frequency (1581 cm$^{-1}$) compared to the *G* band from the monolayer region (1586 cm$^{-1}$) and is in accord with the expectation for lower doping in the BLG compared to monolayer. The Raman spectra presented in Fig. 2 prove that the splitting of the *G* band due to inhomogeneous doping is indeed unique to the overlapped GBs.

Another interesting occurrence in the Raman spectra from the overlapped GBs is the effect of edge structure. Fig. 3 shows Raman spectra collected from two overlapped GBs with similar widths (~180 nm) but with contrasting features. The GB in Fig. 3a exhibits a pronounced *D* band, whereas the GB in Fig. 3b exhibits a negligible *D* band. Both GBs exhibit split *G* bands with a similar degree of splitting, suggesting that the doping levels are the same. It is known that graphene grains can terminate with edges having either an armchair or zigzag geometry. An ideal armchair edge is expected to exhibit large *D* band intensity, whereas an ideal zigzag edge does not exhibit any *D* band [56]. The occurrence of armchair and zigzag edges could therefore account for the contrasting Raman spectra from two similar GBs shown in Fig. 3.

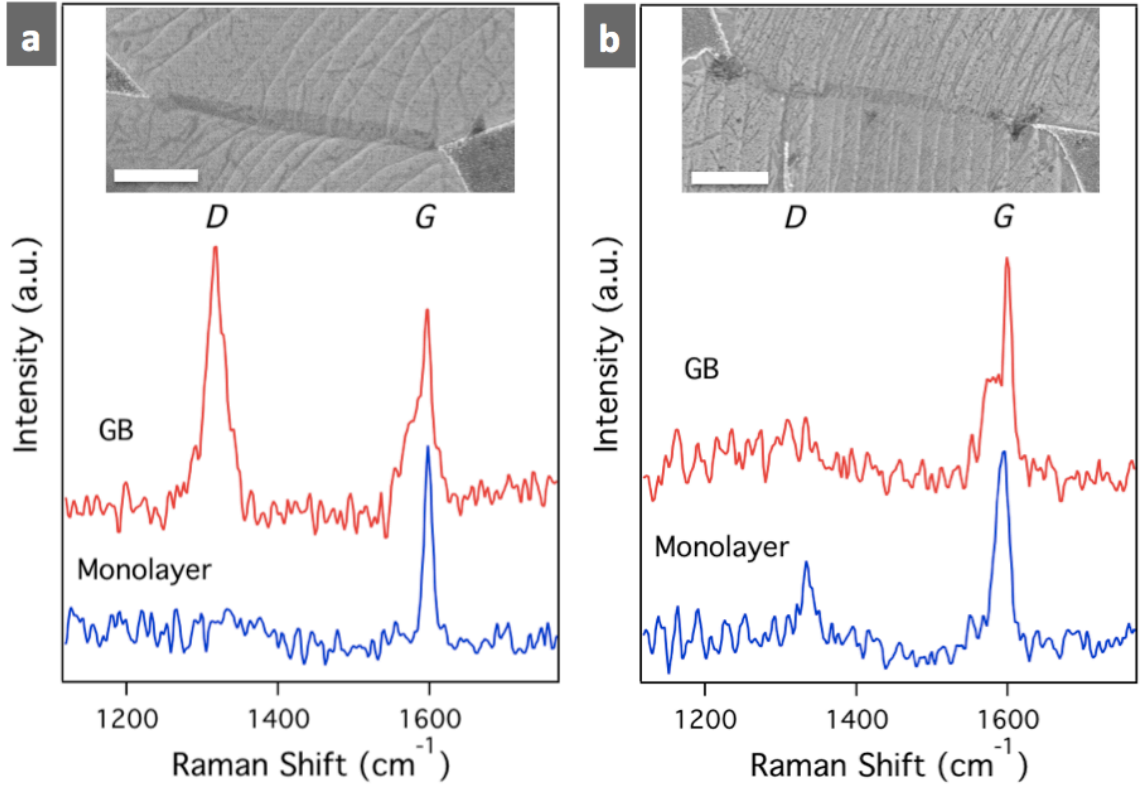

Fig. 3 - Raman spectra collected from two different GB overlaps with similar widths (~180 nm). The overlapped GB in (a) exhibits a pronounced *D* band unlike the GB in (b), suggesting that the GB edge in (a) has an armchair geometry, while the GB in (b) has a zigzag edge. The scale bars are 1 μm.

We now discuss the effect of the width of the GB overlap on their Raman spectra. We found GBs with a wide range of overlap widths ranging from 60 nm to 220 nm in our samples. Remarkably, we find a strong dependence of the degree of splitting in the *G* band on the GB overlap width. Fig. 4a shows the dependence of the difference in peak frequencies between the $G^+$ and $G^-$ bands ($\omega_{G^+} - \omega_{G^-}$) on the GB overlap width. Since the degree of G band splitting is directly related to the charge concentration between the two graphene layers, the positive correlation between ($\omega_{G^+} - \omega_{G^-}$) and the GB overlap width strongly suggests that the degree of doping increases with the width of the GB. The inset in Fig. 4a shows the frequencies of the $G^+$ and $G^-$ peaks as a function of the GB width. The lower frequency $G^-$ peak can be observed to be more strongly dependent on the GB width. The frequency dependence of the $G^+$ and $G^-$ peaks are more clearly shown in Fig. 4b, which plots the difference between the GB and graphene peak frequencies, i.e., $\omega_{G^+} - \omega_G$ or $\Delta\omega_{G^+}$, and $\omega_{G^-} - \omega_G$, or $\Delta\omega_{G^-}$, as a function of the GB overlap width. The $\omega_G$ values correspond to the graphene next to the respective GBs. The decrease in the $G^-$ peak frequency with increasing GB overlap is revealed by the negative slope to the data. On the other hand, the $G^+$ peak blueshifts compared to $\omega_G$ as evinced by the positive slope in the $\Delta\omega_{G^+}$ data. The stronger dependence of $\omega_{G^-}$ on the charge density has been shown previously for inhomogeneously doped BLG [43-45, 47]. The intensity ratio between the $G^+$ and $G^-$ peaks ($I_{G+}/I_{G-}$) also depends strongly on the charge concentration in BLG [43, 44, 47]. As can be seen in Fig. 4c, $I_{G+}/I_{G-}$ increases with the GB overlap width. We note that $I_{G+}/I_{G-}$ does not depend on the length of the GB overlap (inset in Fig. 4c), only the width.

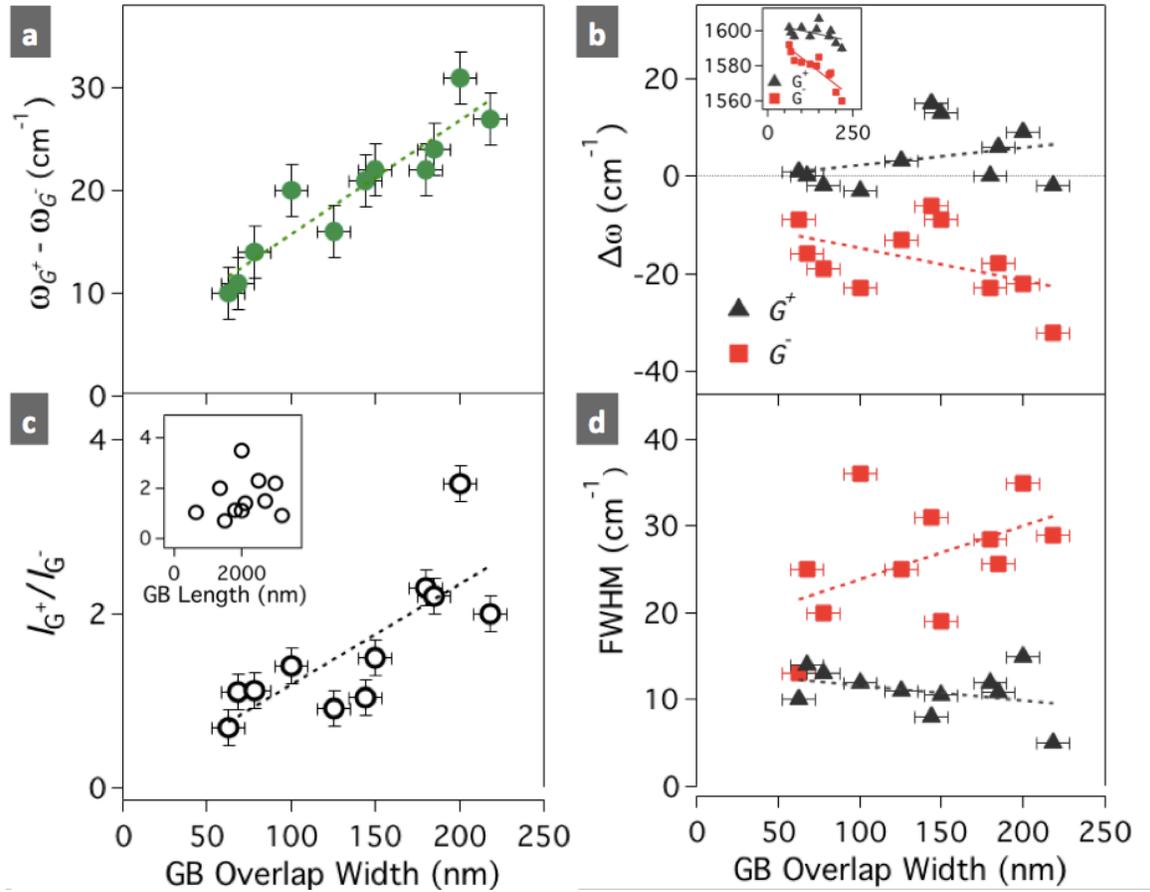

Fig. 4 - Plots of (a) difference between the $G^+$ and $G^-$ peak frequencies, (b) deviation of the $G^+$ and $G^-$ peak frequencies from the G band frequency, (c) Intensity ratios between the $G^+$ and $G^-$ peaks ($I_{G+}/I_{G-}$), and (d) linewidths (FWHM) of the $G^+$ and $G^-$ peaks as a function of the GB overlap width. The inset in (b) shows the peak frequencies of the $G^+$ and $G^-$ peaks as a function of the GB overlap width. The dashed straight line fits in (b), (c), and (d) are guides to the eye.

The different coupling between the symmetric and anti-symmetric modes to electron-hole pairs leads to these changes in the frequencies and intensities of the $G^-$ and $G^+$ peaks. At hole densities greater than $1 \times 10^{13}$ (cm$^{-2}$), i.e. when the Fermi level is greater than the $G$ band phonon energy ($E_F > 0.2$ eV), the $G^-$ peak is dominated by the anti-symmetric mode and is lower in intensity than the $G^+$ peak [44]. We also observe distinct dependencies of the linewidths of the $G^-$ and $G^+$ peaks on the GB overlap width. As shown in Fig. 4d, the $G^-$ and $G^+$ peaks undergo broadening and sharpening, respectively, with increasing GB overlap. Such a contrasting behavior of the $G^-$ and $G^+$ peak linewidths is not expected theoretically. However, it has been seen in BLG samples doped asymmetrically with sulfuric acid [47]. Additionally, broadening of the anti-symmetric mode has been observed due to increased electron-phonon coupling in gated BLG [57, 58]. The dominant anti-symmetric mode at higher charge densities could thus account for the broadening we observe in the $G^-$ peak. Moreover, it is important to note that the previous studies were performed on BLG samples prepared by mechanical exfoliation. In our case the BLG is formed by the overlapping of two individual graphene grains during high temperature CVD growth, and is different from exfoliated BLG. Hence an exact explanation of the observed behavior of optical phonons in the overlapped GBs would require a careful study on their interaction with charges in a controlled (for example, gated) environment.

In order to visualize the charge distribution across the overlapped GBs we perform conductive AFM across the overlapped GBs. Fig. 5a shows an SEM image of an overlapped grain boundary. Figs. 5b and 5c show the corresponding current and friction scans taken from the same area. Although the scans are noisy due to the roughness imparted by the oxidized Cu foil, a clear change in contrast can be observed in the area corresponding to the overlapped GB in both the current and friction scans. Moreover, the linescans across the topography and friction maps show clear differences on and off the overlapped GB (Fig. S7).

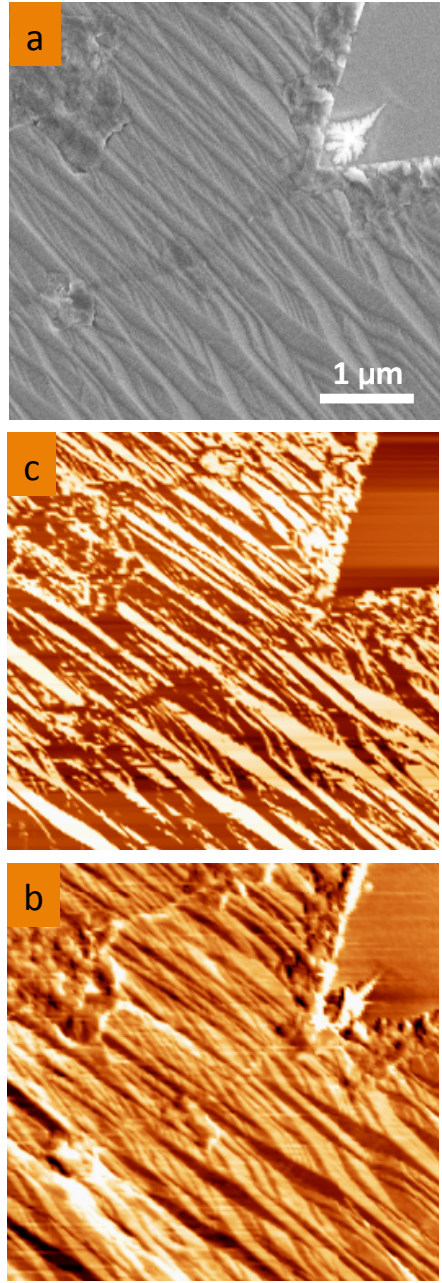

Fig. 5 – (a) SEM image of an overlapped GB and corresponding (b) current and (c) friction maps generated with conductive AFM. The vertical scales in (b) and (c) range from 0–10 nA and 0-10 mv, respectively.

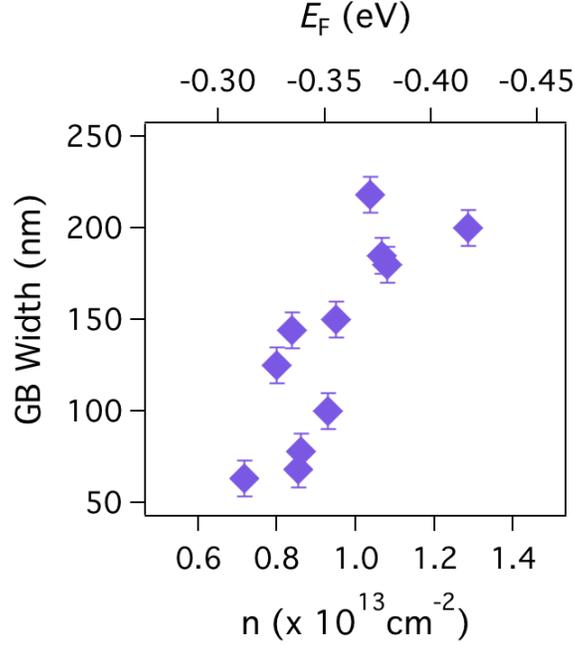

Fig. 6 - Plot of GB overlap width versus estimated charge densities (bottom axis) and calculated Fermi level energies (top axis).

Since the $I_{G+}/I_{G-}$ ratio is expected to depend strongly on the charge concentration theoretically [43, 44], and is also observed experimentally (Fig. 4c and Ref. 47), we compare our $I_{G+}/I_{G-}$ ratios with the previously published values and extract the corresponding charge densities ($n$). The densities increase with the width of the GB overlap from ~0.7 to 1.3 x $10^{13}$ cm$^{-2}$. We also estimate the Fermi level corresponding to these charge densities according to the formula $E_F = -hv_F(n/4\pi)^{1/2}$, where $v_F$ is the Fermi velocity = 1.1 x $10^6$ m$^2$/s [59, 60]. The charge densities and calculated Fermi energies are plotted as a function of the GB overlap width in Fig. 6. Due to hole-doping we find Fermi levels ranging from -0.3 to -0.42 eV below the charge neutrality point as the GB overlap width increases from 60 to 220 nm. Oxygen intercalation in BLG has been previously observed to induce p-type doping with charge densities similar to those observed in the present study [29, 61].

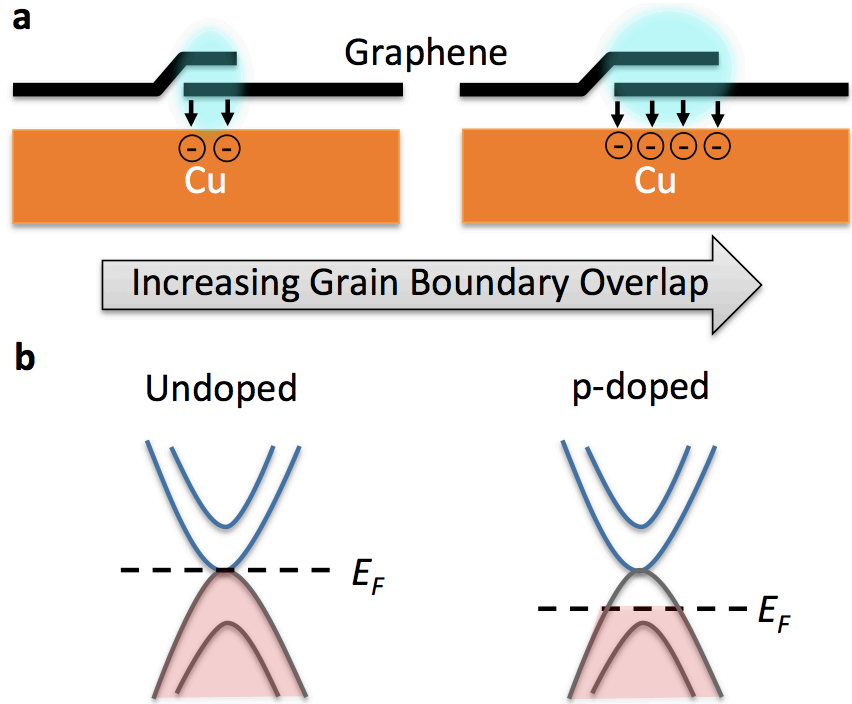

Fig. 6 – (a) Schematic showing inhomogeneous doping across the overlapped GBs with increasing width. The top and bottom layers are doped inhomogeneously from the oxidized Cu substrate. (b) Schematic band diagrams showing the location of the Fermi levels in undoped and p-doped BLG.

The splitting of the *G* band indicates that the bottom and top layers in the GB are doped differently. A schematic showing the inhomogeneous doping of the GB by the copper oxide substrate is shown in Fig. 7a. Increasing overlap causes greater doping of the bottom layer compared to the top graphene layer, thus enhancing the inhomogeneous doping of the GB. As a result the Fermi level drops below the neutrality point due to doping of the GB with holes, as shown in the schematic band diagrams in Fig. 7b. Our study has shown the presence of AB-stacked bi-layer overlapped GBs between merged graphene grains grown by CVD. Raman spectra collected from several GBs indicates inhomogeneous hole doping from the oxidized Cu substrate. The inhomogeneous doping increases with the width of the overlap, thereby tuning the Fermi level of the GB between -0.3 and -0.42 eV. While thorough electrical transport measurements across several GBs would be useful, the observed increase in carrier densities across the AB-stacked overlapped GBs suggest continuous electrical conductivity across overlapped GBs in graphene samples. Furthermore, overlapped GBs with tunable widths and edge structures offer the opporunity to study the performance of BLG nanoribbons [25], as well as unique device geometries such as nanoribbon diodes with negative differential resistance [26] and electromechanical switches based on overlapping nanoribbons [27].

## 4. Conclusions

We have utilized Raman spectroscopy to study overlapped GBs in various graphene grains grown by atmospheric pressure CVD. Raman spectra reveal the overlapped regions to be AB-stacked BLG. Oxidation of the Cu substrate was used as an indirect means to probe the charge environment of the overlapped GBs, which exhibits split $G$ band due to inhomogeneous doping. The splitting of the $G$ band frequencies increases with the width of the overlap, due to the dropping of the Fermi level below the charge neutrality point. The increase in charge carriers in the GBs suggests enhanced and continuous electrical conductivities across graphene samples with AB-stacked overlapped GBs. Furthermore, engineered GBs with tunable widths and edge structure offer promising avenues for their use as nanoribbons in electronic devices.

**TOC Graphic**

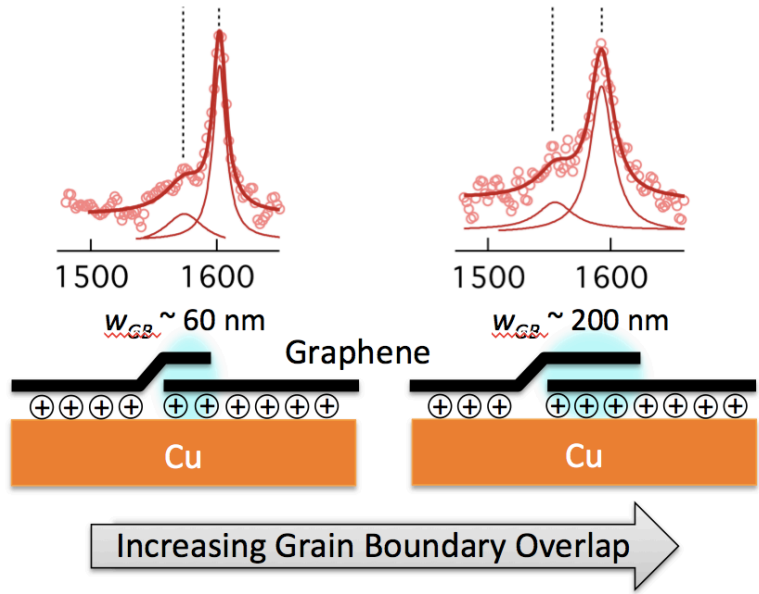